\documentclass[12pt]{article}

\usepackage{graphics,subfigure}
\usepackage{amssymb,amsthm}
\usepackage[driver]{graphicx}

\newcommand{\C}{\mathbb{C}}
\newcommand{\N}{\mathbb{N}}
\newcommand{\R}{\mathbb{R}}

\newcommand{\rme}{{\rm e}}
\newcommand{\rmi}{{\rm i}}

\newcommand{\OO}{{\cal O}}

\newtheorem{claim}{Claim}
\newtheorem{theorem}[claim]{Theorem}

\newtheorem{lemma}[claim]{Lemma}

\begin{document}

\vspace{0.5cm}
\begin{center}
{\bf \Large Approximation by point potentials\\
 in a magnetic field}
\end{center}

\begin{center}
{\large Kate\v{r}ina O\v{z}anov\'a}
\end{center}
\vspace{0.5cm}

\begin{center}
\begin{minipage}{0.85 \textwidth}
{\small \em Department of Mathematical Sciences, 
Chalmers University of Technology, 412 96 G\"{o}teborg, Sweden \\
\rm nemco@math.chalmers.se}
\end{minipage}
\end{center}

\vspace{0.5cm}

\begin{abstract}
We discuss magnetic Schr\"{o}dinger operators perturbed by measures
from the generalized Kato class. Using an explicit Krein-like formula for
their resolvent, we prove that these operators can be approximated in
the strong resolvent sense by magnetic Schr\"{o}dinger operators with
point potentials. Since the spectral problem of the latter operators
is solvable, one in fact gets an alternative way to calculate discrete spectra;
we illustrate it by numerical calculations in the case when the potential 
is supported by a circle.
\end{abstract}

\section{Introduction}

Schr\"{o}dinger operators are used for modelling a particle confined in a
quantum-mechanical system. Depending on a potential which describes how
the particle interacts with its environment, one can consider a wide range 
of physical situations. In this paper, we are particularly interested
in potentials in dimension two supported by zero measure sets; the
supports could be for example graphs, curves or points. 

The motivation to study such operators is based on the fact that they represent
simple mathematical models of various nano-structures like quantum wires, 
photonic crystals, quantum dots, etc. One possible way to describe them is 
via quantum graphs; it means that one considers ordinary differential equations
on the graph edges, which are coupled through boundary conditions at the
graph vertices so that the resulting operator is self-adjoint, see \cite{K}
or \cite{KS}. The operators we are going to deal with yield an alternative 
approach. The particle is not confined to the graph, but it moves in its
vicinity if the potentials are attractive. Hence the latter model is in a sense
more realistic and it enables us to take the tunnelling effect into account.

We aim to prove a limit relation between two classes of operators in
$L^2(\R^2)$ in the presence of a magnetic field: those with attractive 
potentials supported by a curve or a graph on one side, and operators with 
point potentials on the other side. The crucial feature of 
the latter operators is the solvability of their spectral problem; if the 
number of potentials is finite, then the essential spectrum stays unchanged 
and the discrete spectrum can be calculated numerically by solving 
an implicit equation. Therefore, if we are able to find a sequence of point 
potential operators which approximates the given operator, we get an 
approximate method to calculate its discrete spectrum. 

In fact, we will show that the approximation works for a larger family
of potentials than just the ones supported by a curve. The regular
potentials with Kato property will be also included; that is why we speak 
about a generalized Kato class and potentials are replaced by more general 
measures. For a given measure $m$ from the generalized
Kato class, it is possible to define the operator $-\Delta + m$ via
association with a closed and semi-bounded quadratic form, see e.g. 
\cite{BEKS} and \cite{SV}. The second way to define Schr\"{o}dinger operators 
with potentials supported by zero measure sets is by prescribing the 
operator domains, see \cite{Po1} for general singular perturbations, and 
\cite{AGHH} or \cite{GHS} for point potentials; there the domains are given 
by imposing a boundary condition on wavefunctions. 

It was shown in \cite{BFT} that the free Laplacian perturbed by a measure 
with Kato property can be approximated by point potential operators;
in dimension one the convergence is in the norm resolvent sense, while in
dimension three it is in the strong resolvent sense. According to \cite{EN},
the situation in dimension two is similar to the three-dimensional case,
moreover, the authors presented several physical systems where the approximation 
is useful in spectral calculations.

The present task is to prove that the approximation also works in the 
presence of a magnetic field. It turns out 
that the main difficulty is not the proof itself (it easily carries over 
from the non-magnetic case) but rather the lack of information about 
magnetic systems. Namely, we first need to clarify the definition
of perturbations by a measure in section~\ref{sec: perturb}. Then in
section~\ref{sec: krein} we derive an explicit formula for resolvents;
that must be done without using results from \cite{BEKS} directly because 
their proof relies on the positivity preserving property of the free Laplacian. 
Section~\ref{sec: point} deals with point potentials in the presence of a
magnetic field. Finally, in section~\ref{sec: approx} we state the main 
approximation claim and we apply the approximation to a simple example in
section~\ref{sec: appl}, where the magnetic field is homogeneous and the 
potential is supported a circle.

\section{Magnetic Schr\"odinger operator in $\R^2$}

The free magnetic Schr\"{o}dinger operator on $L^2(\R^2)$ is given by
$$
(-\rmi \nabla - A(x))^2,
$$
where $A(x)$ is a vector potential, whose components $A_1$ and $A_2$ belong
to $C^{\infty}(\R^2)$. According to \cite[chapter 1.3]{CFKS}, there exists a 
closed and positive quadratic form $h$,
  \begin{eqnarray}
D(h) &=& \left\{ \psi \in L^2(\R^2) : \; (\partial_j -iA_j)\psi \in L^2(\R^2), \;
j=1,2 \right\} \nonumber
\\
h(\varphi,\psi) &=& \sum_{j=1}^2 ((\partial_j-iA_j)\varphi,(\partial_j-iA_j)\psi
)_{L^2(\R^2)}. \nonumber
  \end{eqnarray}
We define the free magnetic Schr\"odinger operator $H_0$ as the unique 
self-adjoint operator associated with the form $h$, i.e.
  \begin{eqnarray}
D(H_0) &\subset& D(h) \nonumber
\\
(H_0\varphi,\psi) &=& h(\varphi,\psi) \qquad \varphi,\psi \in D(H_0). \nonumber
  \end{eqnarray}
Moreover, from \cite[theorem 1.13]{CFKS} we know that $C^\infty_0(\R^2)$ is a
form core of $H_0$.

By \cite[theorem~14]{BGP1} the resolvent
$(H_0-z)^{-1}$ has an integral kernel $G_0(x,y;z)$ which is continuous
away from the diagonal $x=y$. The singularity of $G_0(x,y;z)$ on the diagonal
is of the same type as the one for the non-magnetic Green function;
following \cite[theorem 15]{BGP2}, it can be rewritten as
$$
G_0(x,y;z) = -{1 \over 2\pi} \, \ln(|x-y|) + G_0^{\mathrm{ren}}(x,y;z),
$$
where $G_0^{\mathrm{ren}}(\cdot;z)$ is continuous on $\R^2 \times \R^2$.
Hence it is possible to introduce the regularized Green function
  \begin{equation}\label{xi}
\xi(a;z) = \lim_{|x-a| \to 0} \left[G_0(x,a;z) + {1 \over 2\pi}\, \ln(|x-y|) 
\right] = G_0^{\mathrm{ren}}(a,a;z).
  \end{equation}
We will need this function when defining a perturbation by point potentials.

In the special case of a homogeneous magnetic field $B$ one can write Green 
function $G_0(z)$ explicitly. For example, in the symmetric gauge,
$A(x)=(-{1 \over 2}B x_2, {1 \over 2}B x_1)$, $B \in \R$, Green function
has the following form, see \cite{DMM}
\begin{equation}\label{homG}
G_0(x,y;z) = {1\over 4\pi}\, \Phi_B(x,y)\, \Gamma \left( {|B|-z \over 2|B|} \right)
U \left( {|B|-z \over 2|B|},1; {|B|\over 2}|x-y|^2 \right),
\end{equation}
$U$ is the irregular confluent hypergeometric function \cite[13.1.33]{AS} and
$\Phi_B$ is a phase factor
$$
\Phi_B(x,y) = \exp \left[ -{\rmi B \over 2} (x_1 y_2-x_2 y_1) - {|B| \over 4}
|x-y|^2 \right].
$$

\section{Perturbation by a measure}\label{sec: perturb}

Next, we perturb the magnetic Hamiltonian $H_0$ by a measure $-\gamma m$ in
the following way,
$$
H_{\gamma m} = H_0 - \gamma m,
$$
where $m$ is a finite positive measure from generalized Kato class, which 
means in dimension two that it satisfies
$$
\lim_{\varepsilon \to 0} \sup_{x \in \R^2} \int_{B(x,\varepsilon)}
|\ln(|x-y|)|\, m(dy) = 0,
$$
with $B(x,\varepsilon)$ denoting the circle of radius $\varepsilon$ centred
at $x$. $\gamma$ is a bounded and continuous function mapping $\Gamma:=
\mathrm{supp}(m)$ into $\R_+$, thus we consider only attractive potentials. 
An example of such measure is the Dirac measure, supported by a 
curve or graph; one can easily check that the condition above holds. 

By \cite[theorem 3.1]{SV} the potential generated by $m$ is 
($-\Delta$)-form bounded with infinitesimally small relative bound. In order 
to define $H_{\gamma m}$ properly, we need a similar form-boundedness
with respect to $H_0$. We cannot use the mentioned result directly as it was
formulated only for Dirichlet quadratic forms, i.e. the ones which are
positivity preserving. Instead, we can employ the diamagnetic inequality to
pass from the non-magnetic system to the magnetic one. In the following, 
$\| \cdot \|_{p,q}$ denotes the norm of an operator acting from $L^p(\R^2)$ 
to $L^q(\R^2)$, $1 \leq p,q \leq \infty$.
\begin{lemma}\label{lem:Kato}
Let $m$ be a positive measure from the generalized Kato class w.r.t. $-\Delta$ 
and let $H_0$ be the self-adjoint operator defined above. Then 
for each $a>0$ there exists $b\in \R$ such that the following inequality
holds for any $\psi \in C_0^\infty (\R^2)$
  \begin{equation}\label{SV inequality}
\int_{\R^2} |\psi(x)|^2\, m(dx) \leq a \, h(\psi,\psi) + b \|\psi \|^2_2.
  \end{equation}
\end{lemma}
{\sl Proof: }
Since $m$ belongs to the generalized Kato class w.r.t. $-\Delta$, by
\cite[theorem 3.1]{SV}, the following inequality is fulfilled for any 
$\psi \in C^{\infty}_0(\R^2)$
  \begin{equation}\label{SV ineq 0}
\int_{\R^2} |\psi(x)|^2\, m(dx) \leq \omega \left[\| \nabla \psi \|_2^2 + 
\zeta \| \psi \|_2^2 \right],
  \end{equation}
where according to \cite[remark 1.7(b)]{SV}, $\omega:= \| (-\Delta +
\zeta)^{-1} m\|_{\infty}$ and thus it decays with $\zeta$ growing to $+\infty$. 
It was proved by approximating $m$ by a sequence of non-negative potentials 
$V_n \in L^2(\R^2) \cap L^{\infty}(\R^2)$ such that
$$
\lim _{n \to \infty} \int_{\R^2} |\psi(x)|^2\, V_n(x) \,dx = \int_{\R^2} 
|\psi(x)|^2\, m(dx) \qquad \forall \; \psi \in C_0^{\infty}(\R^2). 
$$
\cite[theorem 2.1]{SV} states that such sequence exists, 
$\| (-\Delta +\zeta)^{-1}V_n\|_{\infty,\infty} \leq \omega$, $n \in \N$ and
inequality (\ref{SV ineq 0}) holds also when $m(dx)$ is replaced by       
$V_n(x)\,dx$. 

Let us write
  \begin{eqnarray}
\int_{\R^2} |\psi(x)|^2\, V_n(x)\, dx 
&=& \| V^{{1 \over 2}}_n (H_0+\zeta)^{-{1\over 2}} (H_0+\zeta)^{{1\over 2}} 
\psi \|_2^2 \nonumber \\ 
&\leq& \| V^{{1 \over 2}}_n (H_0+\zeta)^{-{1\over 2}} \|^2_{2,2}
\left[ h(\psi, \psi) +\zeta(\psi,\psi) \right]. \nonumber
  \end{eqnarray}
Repeating the proof of \cite[theorem 2.5]{AHS}, we make use of the diamagnetic
inequality \cite{HSU} in this form 
$$
\left| \rme{-t H_0} \psi \right| \leq \rme{-t (-\Delta)} |\psi| \qquad t>0
, \; \psi \in L^2(\R^2).
$$
Then from the expression
$$
(H_0+\zeta)^{-{1\over 2}} = {1 \over \Gamma\left( {1 \over 2}\right) }
\int t^{-{1 \over 2}} \rme{-t \zeta} \rme{- t H_0} \; dt
$$
we get $|(H_0+\zeta)^{-{1\over 2}} \psi| \leq (-\Delta+\zeta)^{-{1\over 2}}
|\psi|$, which in turn yields
$$
\| V^{{1 \over 2}}_n (H_0+\zeta)^{-{1\over 2}} \|^2_{2,2} \leq
\| V^{{1 \over 2}}_n (-\Delta+\zeta)^{-{1\over 2}} \|^2_{2,2} \leq
\| V_n^{{1 \over 2}}(-\Delta+\zeta)^{-1} V_n^{{1 \over 2}}\|_{2,2}.
$$
The Stein interpolation theorem \cite[theorem~IX.21]{RS} and the duality 
between $\|\cdot\|_{1,1}$ and $\|\cdot\|_{\infty,\infty}$ imply
$$
\| V^{{1 \over 2}}_n (H_0+\zeta)^{-{1\over 2}} \|^2_{2,2} \leq
\| V_n (-\Delta+\zeta)^{-1} \|_{1,1}^{{1 \over 2}} \|(-\Delta+\zeta)^{-1} 
V_n \|_{\infty,\infty}^{{1 \over 2}} \leq \omega.
$$
Finally, the convergence of $V_n$ to $m$ and the fact that $\omega \to 0$ 
as $\zeta \to \infty$ finish the proof.
\qed

Since $C_0^\infty (\R^2)$ is dense in $D(h)$ it is possible to define 
linear operator $I_m$
  \begin{eqnarray*}
&& I_m \, : \; D(h) \mapsto L^2(m):=L^2(\R^2,m)
\\
&& I_m \psi = \psi \qquad \forall \psi \in C_0^\infty (\R^2).
  \end{eqnarray*}
Then (\ref{SV inequality}) can be extended to whole $D(h)$ with function 
$\psi$ on the lhs being replaced by $I_m\psi$ and thus $I_m$ is bounded.
Now, consider quadratic form $h_{\gamma m}$ given by
  \begin{eqnarray*}
D(h_{\gamma m}) &=& D(h)
\\
h_{\gamma m} (\psi,\varphi) &=& \int_{\R^2} (\nabla\bar{\psi}(x)+ \rmi A(x)\,
\bar{\psi} (x)).(\nabla \varphi(x) - \rmi A(x)\,\varphi (x))\, dx
\\
&& - \int_{\R^2} I_m \bar{\psi}(x)\, I_m \varphi(x)\, \gamma(x)\, m(dx),
  \end{eqnarray*}
We employ the KLMN theorem, see \cite[theorem X.17]{RS},
to conclude that $h_{\gamma m}$ is lower semi-bounded and closed.
Thus there exists a unique self-adjoint operator $H_{\gamma m}$
associated with this form.

The definition that we have presented above applies to both regular potentials
$m(dx)=V(x)dx$ and potentials supported by zero measure sets $\Gamma$. In 
the latter case there is an alternative way to define operator $H_{\gamma m}$ 
via boundary conditions. Consider an operator which behaves as $H_0$ away from 
the set $\Gamma$
$$
\dot H_{\gamma m} \psi (x) = (-\rmi \nabla - A)^2 \, \psi (x) \qquad x \in \R^2
\setminus \Gamma
$$
with the domain consisting of functions $\psi$ such that $H_0 \psi \in
L^2(\R^2\setminus \Gamma)$, their restriction to $\R^2 \setminus \Gamma$ is
smooth and which are moreover continuous at $\Gamma$
and have a jump in the normal (w.r.t. curve $\Gamma$) derivatives,
$$
{\partial\psi\over\partial n_+}(x) - {\partial\psi\over\partial n_-}
(x) = -\gamma(x)\, \psi(x)\,, \qquad x\in\Gamma\,.
$$
One can check that $\dot H_{\gamma m}$ is e.s.a. and by Green's formula we have
$(\dot H_{\gamma m} \psi, g) = h_{\gamma m} (\psi, g)$ for all $\psi \in D(\dot 
H_{\gamma m})$ and $g \in C^{\infty}_0(\R^2)$. Since $C^{\infty}_0(\R^2)$ is a
core of $h_{\gamma m}$, the closure of $\dot H_{\gamma m}$
can be identified with $H_{\gamma m}$. 
This definition is applicable to curves $\Gamma$ which do not have any 
cusps and only a finite number of smooth edges meet in a node.

\section{Krein-like formula}\label{sec: krein}

Throughout this work, a crucial role will be played by Krein's formula which
gives us an explicit expression for the resolvent of $H_{\gamma m}$. Originally,
the formula was used for Hamiltonians perturbed by a finite number of
point interactions, later, it was generalized to a large family of
operators, see e.g. \cite{Po1}. 

In paper \cite{BEKS} the authors derived the resolvent for the free Laplacian 
perturbed by measure $m$ from generalized Kato class, using the positivity 
property of Green function of the Laplacian. Although their proof does not 
apply to magnetic systems, one would still expect that the resolvent 
corresponding to the operator $H_0$ with perturbed by $m$ should look 
the same in the presence of a magnetic field:
  \begin{equation}\label{krein}
R(z) = R_0(z) + R_{dx,m}(z) \left[{1 \over \gamma} -R_{m,m}(z)
\right]^{-1} R_{m,dx}(z),
  \end{equation}
where $R_{\mu,\nu}(z)$ is an integral operator acting from $L^2(\R^2,\nu)$
to $L^2(\R^2,\mu)$, $\mu$ and $\nu$ are two arbitrary positive Radon measures 
and
$$
R_{\mu,\nu}(z)\psi(x) = \int G_0(x,y;z)\, \psi(y)\, \nu(dy) \qquad 
\mu-\mathrm{a.e.}
$$
Note that $R_{m,dx}(z)= I_m R_0(z)$.

To prove that (\ref{krein}) is indeed the resolvent of $H_{\gamma m}$,
we first show several auxiliary results. 
\begin{lemma}\label{lem:adjoint}
Assume that measure $m$ is finite, i.e. $\int m(dx) =l_m < \infty$
and $z \in \rho(H_0)$.
Then $R_{dx,m}(\bar{z}) = (R_{m,dx}(z))^*$.
\end{lemma}
{\sl Proof:} For $z \in \rho(H_0)$ the operator $R_{m,dx}(z)$ is bounded
and so is its adjoint. We have to check that in the expression
$$
I = (f,R_{m,dx}(z)\psi)_{L^2(m)} =\int m(dy) \int dx \,\bar{f}(y)\, G_0(y,x;z)\, \psi(x)
$$
one can interchange the order of integration. This is possible if the following
integral is finite
$$
I \leq I_1 := \int \int m(dy)\, dx\, |f(y)|\, |G_0(y,x;z)|\, |\psi(x)|.
$$
By Kato's inequality (see \cite{BGP1}) one has
$$
|G_0(x,y;z)| \leq K(x,y;z) = K(x-y,0;z) \qquad {\mathrm a.e.} \; x,y, \in \R^2
$$
for $z$ sufficiently large negative, $K$ is the Green function of the free
Laplacian $-\Delta$ in $L^2(\R^2)$. Thus $I_1$ is dominated by
$\| \psi \|_{L^2(\R^2)}\, l_m^{1/2}\, \| K(z) \|_{L^2(\R^2)}\,
\| f \|_{L^2(m)}$.  Together with the observation
$G_0(x,y;z)=\overline{G_0(y,x;\bar{z})}$ (which is a consequence of $G_0(z)$ 
being a Carleman kernel, see \cite[theorem~16]{BGP1} and \cite[theorem~1]{GMC})
it implies the claim for large negative $z$. For any other $z \in \varrho(H_0)$,
it follows from the first resolvent formula
  \begin{eqnarray*}
R_{dx,m}(z) = R_{dx,m}(z_0) + (z-z_0) R_0(z) R_{dx,m}(z_0). 
  \end{eqnarray*}
\qed

\begin{lemma}\label{lem: adjoint Im}
Consider mappings $I_m$ and $R_{m,dx}(z)$ defined as above and let $z_0 <0$ 
be such that
$$
(\psi,\varphi)_{z_0}:= h(\psi,\varphi) - (z_0 \psi,\varphi)
$$
is an inner product in $D(h)$. (Recall that $h$ is lower semi-bounded and
closed form, thus $D(h)$ with $(\cdot,\cdot)_{z_0}$ is a Hilbert space.)
Then $I_m^* = (R_{m,dx}(z_0))^*$.
\end{lemma}
\noindent The lemma was proved in \cite{B}, let us present its proof for
the sake of completeness. 

{\sl Proof:}
As both operators $I_m$ and $R_{m,dx}(z_0)$ are bounded, their adjoint
operators are bounded, too. For any $f \in L^2(m)$ and $\psi \in L^2(\R^2)$
we have
  \begin{eqnarray*}
(\psi,I_m^* f)_{L^2(\R^2)} &=& h(R_0(z_0)\psi, I_m^* f) - (z_0 R_0(z_0) \psi,
I_m^*f)_{L^2(\R^2)} \\
&=& (R_0(z_0)\psi, I_m^* f)_{z_0} = (I_m R_0(z_0) \psi, f)_{L^2(m)} \\
&=& (R_{m,dx}(z_0)\psi, f)_{L^2(m)} = (\psi, (R_{m,dx}(z_0))^* f)_{L^2(\R^2)}.
  \end{eqnarray*}
In the first line, we have used the fact that $R_0(z_0)$ is the resolvent
corresponding to Hamiltonian $H_0$, associated with $h$. In the second line,
we have simply employed the definition of adjoint operator $I_m^*$; since $I_m$
maps from $D(h)$ with the inner product $(\cdot, \cdot)_{z_0}$ to $L^2(m)$ it
reads $(\phi, I^*_m f)_{z_0} = (I_m \phi, f)_{L^2(m)}$, $\phi \in D(h)$ and 
$f \in L^2(m)$. \qed
\begin{lemma}\label{lem: R_dx_m}
Assume $z \in \rho(H_0)$ and $f \in L^2(m)$.
Then $R_{dx,m}(z) f \in D(h)$ and
$$
h(R_{dx,m}(z)f, \psi) - (z R_{dx,m}(z) f, \psi)_{L^2(\R^2)} = (f, I_m
\psi)_{L^2(m)}
$$
for all $\psi \in D(h)$.
\end{lemma}
{\sl Proof:} Using the first resolvent formula, one has
$$
R_{dx,m}(z) f = R_{dx,m}(z_0) f + (z-z_0) R_0(z) R_{dx,m}(z_0) f,
$$
where $z_0$ is the same as in the previous lemma; then the first
term equals $I_m^* f$ and hence it belongs to $D(h)$. Also the second term
belongs to $D(h)$ as the free resolvent $R_0(z)$ maps to $D(H_0) \subset D(h)$.

To prove the second claim we substitute $R_{dx,m}(z)f$ from the above formula
and we use lemma~\ref{lem: adjoint Im},
  \begin{eqnarray*}
&& h(R_{dx,m}(z)f, \psi) - (z R_{dx,m}(z) f, \psi)_{L^2(\R^2)} =
\phantom{AAAAAAAAAAAAAAAAA}
\\ &&\phantom{A} = (R_{dx,m}(z_0) f, \psi)_{z_0} -((z-z_0)R_{dx,m}(z_0)f,
\psi)_{L^2(\R^2)} \\
&&\phantom{A=} + h((z-z_0)R_0(z)R_{dx,m}(z_0)f,\psi) - (z(z-z_0) R_0(z)
R_{dx,m}(z_0)f, \psi)_{L^2(\R^2)} \\
&&\phantom{A} = (I_m^* f, \psi)_{z_0} - ((z-z_0)R_{dx,m}(z_0)f, \psi)_{L^2(\R^2)}
\\ &&\phantom{A=}+ ((z-z_0)R_{dx,m}(z_0)f,\psi)_{L^2(\R^2)} \\
&&\phantom{A} = (f, I_m \psi)_{L^2(m)}. 
\phantom{AAAAAAAAAAAAAAAAAAAAAAAAAAAAAAAA} \qed
  \end{eqnarray*}

  \begin{theorem}\label{thm: krein}
Suppose that $1 / \gamma - R_{m,m}(z)$ is invertible. Then $R(z)$ 
given by (\ref{krein}) is defined on $L^2(\R^2)$ and it is the resolvent of
$H_{\gamma m}$.
  \end{theorem}
{\sl Proof:} Take arbitrary $\psi \in L^2(\R^2)$ and $\varphi \in D(h)$, then
by lemma~\ref{lem: R_dx_m} and Krein's formula (\ref{krein}) $R(z)\psi$
belongs to $D(h)$. We have to check that
$$
I := h_{\gamma m} (R(z)\psi, \varphi) - (zR(z)\psi, \varphi)_{L^2(\R^2)} =
(\psi, \varphi)_{L^2(\R^2)}.
$$
Denote $g=(1 /\gamma -R_{m,m}(z))^{-1} R_{m,dx}(z)\psi$, then
  \begin{eqnarray*}
I &=& h(R_0(z)\psi, \varphi) - (z R_0(z)\psi, \varphi)_{L^2(\R^2)}
- (I_m R_0(z)\psi, \gamma I_m \varphi)_{L^2(m)} \\
&+& h(R_{dx,m}(z)g, \varphi) - (z R_{dx,m}(z) g, \varphi)_{L^2(\R^2)}
- (I_m R_{dx,m}(z)g, \gamma I_m \varphi)_{L^2(m)}
  \end{eqnarray*}
According to lemma~\ref{lem: R_dx_m}, the fourth and fifth term give
together $(g,I_m \varphi)_{L^2(m)}$, so one gets
$$
I = (\psi, \varphi)_{L^2(\R^2)} + \Big(-R_{m,dx}(z)\,\psi +{1 \over \gamma}
\,g - R_{m,m}(z)\, g, \gamma\, I_m \varphi \Big)_{L^2(m)},
$$
finally, employing the definition of $g$ shows that the second term
equals zero. \qed

One can fulfil the hypothesis that the operator $1 / \gamma -R_{m,m}(z)$ 
is invertible easily by choosing sufficiently large negative $z$; it follows 
from the next lemma, see \cite[corollary~2.2]{BEKS}.
In the following, $\| T \|_{p,q}$ denotes the norm of an operator $T$ 
acting from $L^p(m)$ to $L^q(m)$.
\begin{lemma}\label{lem: norm of R}
There exists $\tilde{z} < 0$ such that $\| \gamma R_{m,m}(z) \|_{2,2} <1$ for
all $z < \tilde{z}$.
\end{lemma}

{\sl Proof:} Since the measure $m$ belongs to Kato class and $\gamma$
is bounded, we can find $0<a<1$ and $0<b<\infty$ such that
$$
\int_{\R^2} |I_m \psi(x)|^2 (1+\gamma(x)^2)\, m(dx) \leq a\, h(\psi,\psi)
+ b\, (\psi,\psi)_{L^2(\R^2)}
$$
for all $\psi \in D(h)$.
We put $\tilde{z}=-b/a$, the rhs of the inequality then reads
$$
a \, (\psi,\psi)_{\tilde{z}} \,:=\, a\, h(\psi,\psi) - a\, (\tilde{z} 
\psi, \psi)_{L^2(\R^2)}.
$$
Next, we take any $f \in L^2(m)$ and $z < \tilde{z}$ and
introduce a set
$$
S_z := \{\psi \in D(h): \; (\psi,\psi)_z =1 \}.
$$
Consequently, we have
  \begin{eqnarray*}
&&\int_{\R^2} |I_m R_{dx,m}(z) f(x)|^2\, \gamma^2(x)\, m(dx) \leq a \,
(R_{dx,m}(z)f,R_{dx,m}(z)f)_z \phantom{AAAAA}  \\
&&\phantom{A} \leq a \sup_{\psi \in S_z} | (R_{dx,m}(z) f,
\psi)_z|^2 = a \sup_{\psi \in S_z} | (f, I_m\psi)_{L^2(m)}|^2 \\
&&\phantom{A} \leq a \int |f(x)|^2\, m(dx) \sup_{\psi \in S_z} \int
|I_m \psi(x)|^2\, m(dx) \\
&&\phantom{A} \leq a\, \|f\|^2_{L^2(m)}\, a\, (\psi,\psi)_z = a^2\, \|f\|^2_{L^2(m)}
\phantom{AAAAAAAAAAAAAAAAAAAA} \qed
  \end{eqnarray*} 

For further use we rewrite the Hamiltonian $H_{\gamma m}$ into the
form $H_0 -{1 \over \alpha}\mu$, where
$$
\mu = {\gamma m \over \int \gamma m}, \qquad \alpha = {1 \over \int
\gamma m}.
$$
Since both function $\gamma$ and measure $m$ are positive, the
coupling constant $\alpha$ and normalized measure $\mu$ are positive, too.
The resolvent $R(z) = (H_{\gamma m}-z)^{-1}$ then acts on arbitrary $\psi \in
L^2(\R^2)$ as follows
  \begin{equation}\label{krein mu}
R(z) \psi = R_0(z) \psi + R_{dx,\mu}(z) [\alpha -R_{\mu,\mu}(z)]^{-1} 
R_{\mu,dx}(z) \psi.
  \end{equation}
According to lemma~\ref{lem: norm of R}, 
$\|{1\over\alpha} R_{\mu,\mu}(z)\|_{2,2}$ is less than 1 for sufficiently 
large negative $z$, one can prove the same about the norm 
$\|\cdot\|_{\infty,\infty}$ using the Kato property of $\mu$. From now on 
we consider only large negative $z$ so that both norms above are less than 1.
The second term on the rhs of (\ref{krein mu}) can be substituted by $R_{dx,\mu}
(z)\sigma$, where $\sigma \in L^2(\mu)$ is the unique solution to the equation
  \begin{equation}\label{def sigma}
[\alpha - R_{\mu,\mu}(z)]\sigma = R_{\mu,dx}(z) \psi \qquad \mu-\mathrm{a.e.}
  \end{equation}
By \cite[theorem 16]{BGP1} domain $D(H_0)$ is embedded into the space of 
continuous and bounded function on $\R^2$, thus $R_0(z)\psi$ is bounded and 
continuous in $\R^2$ and the same is of course true for function 
$R_{\mu,dx}(z) \psi$ on $\Gamma$. Adding the information about norms of 
${1\over \alpha} R_{\mu,\mu}(z)$, we may conclude that $\sigma$ is bounded 
and continuous on $\Gamma$ as well.

\section{Point potentials}\label{sec: point}
Next consider a magnetic Schr\"{o}dinger operator $H_{Y,\alpha}$ with finitely
many point potentials placed at points $a \in Y \subset \Gamma$, $|Y|$
denotes the number of potentials. The operator is defined via self-adjoint extensions;
away from the points from $Y$ it behaves as the free operator $H_0$ and the
wavefunctions from its domain must have following behaviour in the vicinity of
each point $a \in Y$,
\begin{equation}\label{bc}
\psi(x) = \ln|x-a| L_0(\psi,a) + L_1(\psi,a)+\OO(|x-a|)
\end{equation}
with coefficients $L_0$ and $L_1$ fulfilling the boundary condition
$$
L_1(\psi,a) + 2\pi \alpha(a) L_0(\psi,a) = 0 \qquad \forall a \in Y.
$$
For further details concerning point potentials see e.g. \cite{AGHH} for
the non-magnetic case and \cite{GHS} for the magnetic case. In general, one 
can choose any real number $\alpha(a)$ for each potential independently, here,
we make a special choice $\alpha(a)=\alpha|Y|$ for all $a \in Y$, with $\alpha$
defined in the previous subsection.

The resolvent $(H_{Y,\alpha}-z)^{-1}$ is given by Krein's formula,
  \begin{equation}\label{krein point}
(H_{Y,\alpha}-z)^{-1}\psi \,(x) = R_0(z)\psi \,(x) + \sum_{y,y' \in Y}
[\Lambda_{Y,\alpha}(z)]^{-1}(y,y') \,G_0(x,y;z) \, R_0(z)\psi \,(y'),
  \end{equation}
where $\Lambda_{Y,\alpha}(z)$ is a matrix $|Y|\times |Y|$,
  \begin{eqnarray*}
\Lambda_{yy'} = \left\{ \begin{array}{ll} |Y|\alpha - \xi(y;z)
 \; & y = y'
\\ - G_0(y,y';z) \; & y \neq y',
\end{array} \right.
  \end{eqnarray*}
$\xi(y;z)$ is the regularized Green function (\ref{xi}).
For a homogeneous magnetic field $B$, we can calculate $\xi(a;z)$ explicitly;
it does not depend on the potential 
position $y$ and it equals $-{1\over 4\pi} \left[ \psi \left( 
{|B|-z \over 2 |B|} \right) + 2 C_E + \ln \left( {|B|\over 2}\right) \right]$,
with $C_E$ denoting the Euler constant. The second term on the rhs of 
(\ref{krein point}) can be rewritten as $\sum_{y\in Y} G_0(x,y;z) q_y$, 
where $q$ is the $|Y|$-dimensional vector which solves
  \begin{equation}\label{def q}
R_0(z) \psi\, (y) = \sum_{y' \in Y} (\Lambda_{Y,\alpha}(z))(y,y') q_{y'}
\qquad \forall \; y \in Y.
  \end{equation}

One possible way to make matrix $\Lambda$ invertible is to take a 
sufficiently large set $Y$.
\begin{lemma}\label{lem: Schur norm}
Let $Y_n$ be a sequence of subsets of $\Gamma$ with $|Y_n| \to \infty$ as 
$n \to \infty$,
and assume that for some $\tilde{\alpha}<\alpha$ we have
$$
\sup_{n \in \N} {1 \over |Y_n|} \sup_{x \in Y_n} \sum_{y \in Y_n \setminus \{x\}}
|G_0(x,y;z)| \leq \tilde{\alpha} < \alpha.
$$
Then there exist $C>0$ and $n_0 \in \N$ such that $\Lambda_{Y_n,\alpha}(z)$ is
invertible and
$$
 \left\| \left( {1 \over |Y_n|} \Lambda_{Y_n,\alpha}(z) \right)^{-1}
 \right\|_{2,2} < C
$$
for all $n \geq n_0$.
\end{lemma}
\noindent Here and in the following section, the symbol $\| T \|_{p,q}$ denotes 
the operator norm of a matrix $T$, acting from $\ell^p(Y)$ to $\ell^q(Y)$
and correspondingly, $\| \cdot \|_p$ is the norm in $\ell^p(Y)$.

{\sl Proof:} Let us split matrix $\Lambda_{Y_n,\alpha}(z)/|Y_n|$ into
the diagonal and non-diagonal part. On the diagonal, $\xi(\cdot;z)$ stays bounded
because it is a continuous function on a compact set $\Gamma$. Therefore 
the diagonal part behaves as $(\alpha+ \OO(|Y_n|^{-1})) \delta_{yy'}$ 
for large $n$, its norm in $\ell^2(Y_n)$ converges to $\alpha$ and so it is 
invertible. The non-diagonal part, denoted by $R_n$, is given by 
$- {1\over |Y_n|} G_0(y,y',z)(1-\delta_{yy'})$. Using Schur-Holmgren bound 
(see \cite[appendix C]{AGHH}) we arrive at
$$
\| R_n \|_{2,2} \leq {1 \over |Y_n|} \sup_{x \in Y_n} \sum_{y \in Y_n \setminus \{x\}}
|G_0(x,y;z)| \leq \tilde{\alpha} < \alpha,
$$
hence the whole matrix is invertible. \qed

\section{Approximation}\label{sec: approx}

Now we are ready to formulate the main approximation claim. Its proof
follows closely the one in \cite{BFT}.
  \begin{theorem}\label{thm: approximation}
Let $\Gamma$ be a compact and non-empty set in $\R^2$ and let $m$ be a finite
positive measure with $\Gamma:=\mathrm{supp}(m)$, which belongs to the Kato class.
Let $\gamma$ be a bounded and continuous function on $\Gamma$, which attains
only positive values. Consider sufficiently large negative $z$ such that
the equation (\ref{def sigma}) has a unique solution $\sigma$ with a
bounded and continuous version on $\Gamma$.
Suppose further that there is a sequence of sets $Y_n \subset \Gamma$ with
$|Y_n|\to\infty$ as $n \to \infty$ and satisfying following three conditions:
  \begin{equation}\label{hypothesis 1}
{1 \over |Y_n|} \, \sum_{y \in Y_n} f(y) \to \int f(y) \mu(dy)
  \end{equation}
for any bounded and continuous function $f$ on $\Gamma$,
  \begin{equation}\label{hypothesis 2}
\sup_{n \in \N} {1 \over |Y_n|} \sup_{x \in Y_n} \sum_{y \in Y_n \setminus \{x\}}
|G_0(x,y;z)| \leq \tilde{\alpha}
  \end{equation}
for some $\tilde{\alpha} < \alpha$, and finally
  \begin{equation}\label{hypothesis 3}
\sup_{x \in Y_n} \left| {1 \over |Y_n|} \sum_{y \in Y_n
\setminus \{x\} } \sigma(y) G_0 (x,y;z) - (R_{dx,\mu}(z) \sigma)(x)
\,\right| \; \to \; 0
  \end{equation}
for $n \to \infty$.
Then operators $H_{Y_n,\alpha}$ converge to $H_{\gamma m}$
in the strong resolvent sense as $n \to \infty$.
  \end{theorem}

{\sl Proof:} For self-adjoint operators the weak resolvent convergence implies 
the strong resolvent convergence, thus it is enough to prove that
$$ 
I_n := (\phi, (H_{Y_n,\alpha}-z)^{-1}\psi -(H_{\gamma
m}-z)^{-1}\psi )_{L^2(\R^2)} \; \to 0 \qquad \mathrm{as} \qquad n
\to \infty .
$$
Employing the alternative expressions for both resolvents one gets
$$ 
I_n = \left(\phi, \sum_{y \in Y_n} G_0(\cdot,y;z) \,q_y - R_{dx,\mu}(z) \,
\sigma \right)_{L^2(\R^2)}, 
$$
where $q$ and $\sigma$ are given as solutions to equations (\ref{def q}) and
(\ref{def sigma}), respectively.
  \begin{eqnarray*} 
I_n &=& \sum_{y \in Y_n} \overline{R_0(z) \phi}(y)\, q_y - (R_{\mu,dx}(z) 
\phi,\sigma )_{L^2(\mu)} \\
&=& \sum_{y \in Y_n} \overline{R_0(z) \phi}(y) \left(q_y - {1 \over |Y_n|}\,\sigma
(y) \right) \\
&& + \sum_{y \in Y_n} \overline{R_0(z) \phi}(y) \,{1 \over |Y_n|}\,\sigma(y)
- \int \overline{I_{\mu} R_0(z) \phi} (y') \,\sigma(y') \,\mu(dy').
  \end{eqnarray*}
Since $R_0(z)\phi$ has a bounded and continuous version (and it can be 
identified with $I_{\mu} R_0(z)\phi$), by hypothesis (\ref{hypothesis 1}) 
the last two term vanish in the limit and for the first term we only have 
to show that
$$
\left\| {v^{(n)} \over |Y_n|} \right\|_1 \to 0, \qquad v^{(n)}_y:=|Y_n|q_y -
\sigma(y).
$$
Comparing equations (\ref{def q}) and (\ref{def sigma}), one obtains
following expression for $v^{(n)}$,
  \begin{eqnarray*}
\sum_{y' \in Y_n}  {1 \over |Y_n|} (\Lambda_{Y_n,\alpha}(z))(y,y')\, v^{(n)}_{y'} 
&=& {\sigma(y) \over |Y_n|} \, \xi(y;z) 
+ {1 \over |Y_n|} \sum_{y' \neq y} G_0(y,y';z) \,\sigma(y') \\
&-& \int I_{\mu} G_0(y,y';z) \,\sigma(y') \,\mu(dy').
  \end{eqnarray*}
Last two terms on the rhs vanish because of the hypothesis (\ref{hypothesis 3}) and
also the first term goes to zero as $n \to \infty$ since the numerator is
a bounded function of $y$. So when we denote the vector with elements given 
by the rhs as $w^{(n)}$, then the norm $\|w^{(n)} \|_{\infty}$ tends to zero.

By hypothesis (\ref{hypothesis 2}) and lemma~\ref{lem: Schur norm}, matrix 
$\Lambda_{Y_n,\alpha}(z)/|Y_n|$  is invertible and the operator norm of its inverse
in $\ell^2(Y_n)$ is bounded by some $C$. Hence we can write
  \begin{eqnarray*}
{1 \over |Y_n|}\, \| v^{(n)}\|_1 \,\leq\, {1 \over |Y_n|} \left\| \left( 
{1 \over |Y_n|}\, \Lambda_{Y_n,\alpha}(z) \right)^{-1} \right\|_{\infty,1} \, 
\| w^{(n)}\|_{\infty} \,\leq\, {1 \over |Y_n|}\, |Y_n|\, C\,\, \| w^{(n)}\|_{\infty}
  \end{eqnarray*}
since $\|\cdot\|_{\infty,1} \leq |Y_n| \, \|\cdot\|_{2,2}$ for operators on 
$\C^{|Y_n|}$.
\qed

We have formulated the approximation result for the two-dimensional situation,
however, it could be proved also in dimension three, provided several 
modifications are made. First of all, the generalized Kato class is different in
$\R^3$. Measure $m$ belongs to it if
$$
\lim_{\varepsilon \to 0} \sup_{x \in \R^3} \int_{B(x,\varepsilon)}
{1 \over |x-y|}\, m(dy) = 0,
$$
where $B(x,\varepsilon)$ is the sphere of radius $\varepsilon$ centred
at $x$. A potential with a zero measure support fulfils the condition above
only if the codimenion of its support is equal to 1. Therefore one may 
approximate for example the potentials supported by compact surfaces.

The main technical difficulty in dimension three comes from the fact that the 
Green function of a magnetic Schr\"{o}dinger operator has a divergent term 
which depends on the given vector potential, see e.g. \cite[section 5]{BGP2}. 
Correspondingly, the vector potential also enters the definition of point 
potentials, hence the definition has to be modified as in \cite[section~4]{EN-l}.
Moreover, the divergent term is not logarithmic, hence the term $\ln |x-a|$ 
in (\ref{bc}) must be replaced by $1/|x-a|$.

\section{Application}\label{sec: appl}

One motivation for this paper was to obtain an alternative
method to calculate discrete spectra. We have proved that any operator
$H_{\gamma m}$, defined as $H_0$ perturbed by measure $m$ with Kato property, 
can be approximated by point potential Hamiltonians in the strong resolvent 
sense. Hence for each eigenvalue of $H_{\gamma m}$ there exists a sequence 
of eigenvalues of the latter operators converging to it. 

It is very natural to apply the
approximation to a system whose spectral problem is exactly solvable so
that we can compare the exact and approximate eigenvalues, obtained by
numerical calculation. As an essential prerequisite for numerical calculations
is to have an explicit formula for Green function $G_0(z)$, we restrict
the application only to the case of a homogeneous magnetic field $B$. Then
one can employ the expression (\ref{homG}).

To demonstrate how one can use the approximation to calculate discrete
spectra, let us present following example. Suppose that the potential is 
attractive and constant and it is supported by a circle with radius $R$. 
The potential can be thus described by two parameters $R >0$ and $\gamma >0$.
The easiest choice of point potential operators is the following: we 
place $N$ points equidistantly along the circle and put 
$\alpha= N/(2\pi R \gamma)$.
The spectrum of $H_{\gamma m}$ consists of Landau levels
$$
\sigma_{\mathrm{ess}}(H_{\gamma m}) \,=\, \{ |B| (2m+1): \, \, m=0,1,\ldots \}
$$
and eigenvalues which have split off from the Landau levels because of the
presence of the potential. Since we consider the potential being attractive, 
the eigenvalues are below the level they have arisen from (and moving further 
down as the coupling constant $\gamma$ grows.)

To find the eigenvalues explicitly, one has to decompose the operator
into angular momentum subspaces and then to calculate the eigenvalues
numerically by solving an implicit equation in each subspace, see
\cite{ET}. The resulting picture is that there is one sequence of eigenvalues
in each gap between two adjacent Landau level and below the
lowest one, with the limit point at the upper Landau level.

\begin{figure}[!t]
%
\centering
\subfigure[]{
\label{fig:spec1:a}
\includegraphics[angle=-90, width=0.45\textwidth]{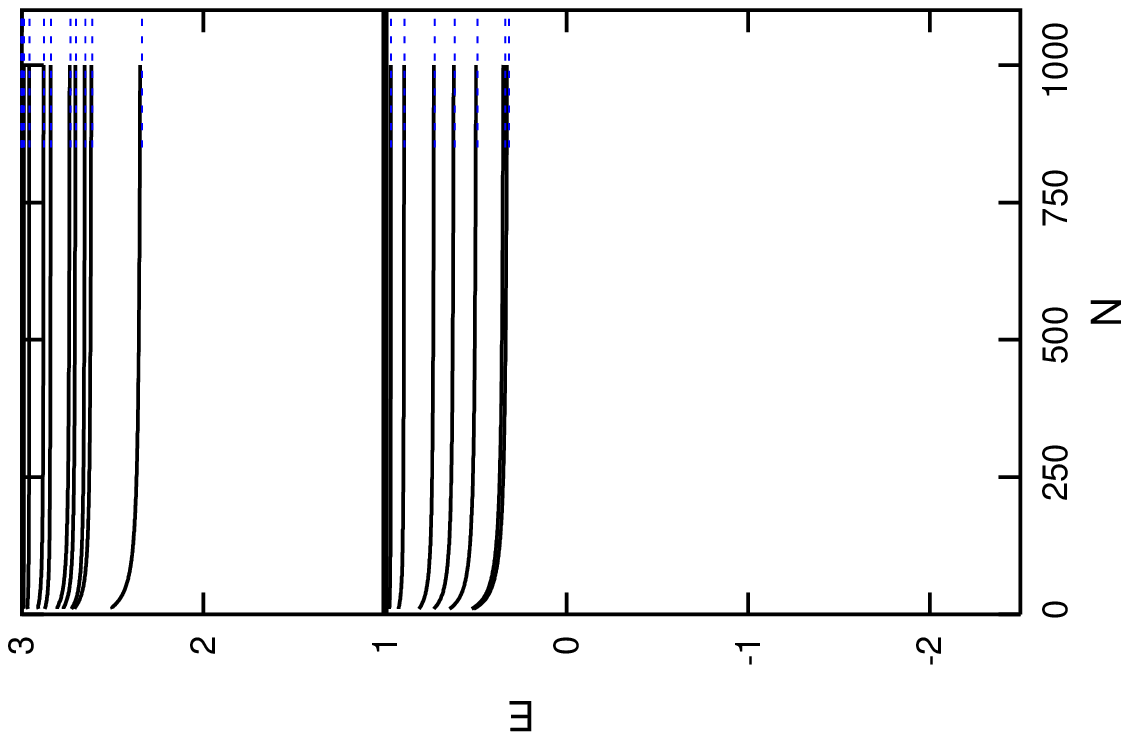}
}
\hfill
\subfigure[]{
\label{fig:spec1:b}
\includegraphics[angle=-90, width=0.45\textwidth]{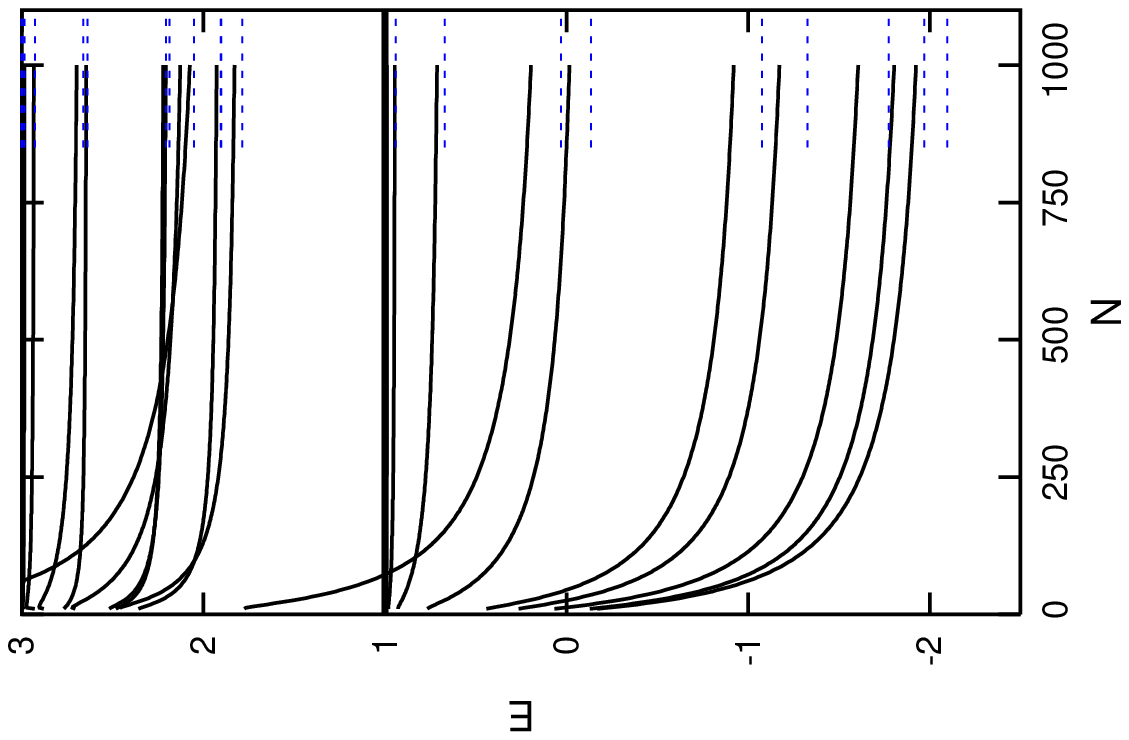}
}
%
%
\caption{The dependence of the approximate eigenvalues on the number 
of point potentials for $B=1$, $R=2$ and $\gamma=1$ (a) and $\gamma=3$ (b).
The dashed lines represent the exact eigenvalues of $H_{\gamma m}$.}
\label{fig:spec1}
\end{figure}

Figure~\ref{fig:spec1} depicts the comparison of the approximate and
exact eigenvalues in two lowest gaps for two situations which differ
only in the coupling constant $\gamma$. Figure \ref{fig:spec1:b}  
corresponds to a stronger attractive potential $\gamma=3$, therefore the 
eigenvalues are further from the Landau levels than those in \ref{fig:spec1:a}
where $\gamma=1$. 
We observe that the approximate eigenvalues tend to the exact ones as
the number of point potentials grows and that the convergence is slower
when the coupling is stronger. One can roughly estimate that the 
convergence rate is of the type $c N^{-a}$, where according to
numerical calculations, $a$ appears to be around $1/2$, while coefficient
$c$ depends strongly on the coupling constant $\gamma$.

A close inspection of the eigenfunctions of the point potential operators
would reveal that they have logarithmic peaks at the potential sites,
since they are given as linear combinations of free Green
functions, see e.g. \cite[chapter II]{AGHH}. 
Although by \cite[theorem~3.4]{Po2}, these wavefunctions yield an 
approximation of wavefunctions of $H_{\gamma m}$, the peaks are of course 
absent in exact eigenfunctions. We believe that this fact is partly 
responsible for the slow convergence of the approximate energies
of the bound states to the exact ones.

All the features we have described were observed in the non-magnetic case,
see \cite{EN}, with the exception that there one deals only with one gap 
$(-\infty,0)$ and the number of eigenvalues in the gap is finite. 
In the absence of magnetic field the ground state of $H_{\gamma m}$
corresponds to angular momentum $l=0$, the remaining bound states correspond
to $\pm l$ and they are double degenerate. In magnetic field there is no
such degeneracy; eigenvalues for angular momenta with opposite signs
are different, because there is an extra angular momentum coming from the
magnetic field. As figure \ref{fig:spec1:b} 
suggests, also the approximate eigenvalues (and in particular, their 
dependence on the number of point potentials) behave differently: 
the eigenvalue crossing the Landau level $B=1$ tends to the eigenvalue of 
$H_{\gamma m}$ for $l=1$, while the eigenvalue for $l=-1$ is the limit 
point of the second lowest approximate eigenvalue. 

\subsection*{Acknowledgement}

The research is supported by the Marie Curie grant MEIF-CT-2004-009256.
The author thanks P. Exner for his constant support throughout the work
and J. F. Brasche for many helpful suggestions. 


\bibliographystyle{plain}

\end{document}